\documentstyle[prl,aps,epsf,float,times]{revtex}
\begin{document}
\draft
\twocolumn[\hsize\textwidth\columnwidth\hsize\csname@twocolumnfalse\endcsname
\title{Incoherent Pair Tunneling as a Probe of the Cuprate Pseudogap}
\author{Boldizs\'ar Jank\'o, Ioan Kosztin and K. Levin} 
\address{The James Franck Institute, The University of Chicago, 
  5640 S. Ellis Avenue, Chicago, Illinois 60637}
\author{M. R. Norman}
\address{Materials Science Division, Argonne National Laboratory, Argonne,
  Illinois 60439} 
\author{Douglas J. Scalapino}
\address{Department of Physics, University of California, Santa Barbara,
  California 93106} 
\date{\today} \maketitle
\begin{abstract}
  We argue that incoherent pair tunneling in a cuprate superconductor
  junction with an optimally doped superconducting and an underdoped
  normal lead can be used to detect the presence of pairing
  correlations in the pseudogap phase of the underdoped lead.  We
  estimate that the junction characteristics most suitable for
  studying the pair tunneling current are close to recently
  manufactured cuprate tunneling devices.
\end{abstract}
] 

The pseudogap -- a depletion of the single particle spectral weight
around the Fermi energy -- is considered to be one of the most
convincing manifestations of the unconventional nature of cuprate
superconductivity\cite{bglevi}. The pseudogap regime sets in as the
temperature is lowered below a crossover temperature $T^*$, and
extends over a wide range of temperatures in {\it underdoped} samples
\cite{Oda}. While the pseudogap is clearly present in the spin channel
\cite{magnetic}, optical conductivity data \cite{optcond} suggest that
the same mechanism is responsible for the gapping of the charge
degrees of freedom as well.  In addition, specific heat data
\cite{loram} also provide evidence that a gap opens below $T^*$. It
has been suggested \cite{2e,Janko,Maly} that precursor superconducting
pairing fluctuations may be responsible for these phenomena; the
observations of a smooth crossover from the pseudo to superconducting
gap seen in angle-resolved photoemission \cite{arpes} and scanning
tunneling spectroscopy \cite{renner} lend support to this idea. There
are, however, several other competing proposals that do not
necessarily involve charge 2e pairing \cite{SNS97}.  It is therefore
of interest to find an experiment which can provide a direct test of
the superconducting precursor scenario.  Here we propose and analyze
an experiment involving incoherent pair tunneling which provides such
a test \cite{SO5}.

The measurement of the pair susceptibility in the normal state of a
superconductor, is in principle similar\cite{djs} to other - say,
magnetic - susceptibility measurements: we are interested in finding
out the linear response of the system to a polarizing external field.
In the present case the role of the external field is played by the
rigid pair field of a second superconductor below its transition
temperature, which couples to the fluctuating pair field of the normal
lead. This coupling leads to an observable contribution to the
tunneling current - the incoherent pair tunneling current - provided
that the normal state has sizeable pairing correlations.

\begin{figure}
\centerline{\epsfxsize=2.2in\epsffile{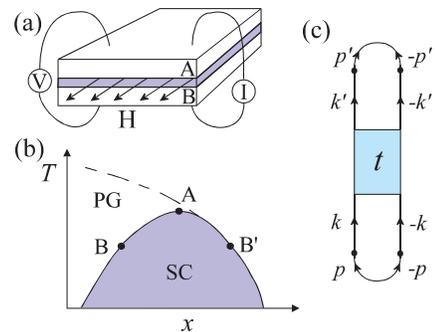}}
\vspace*{1ex}
\caption{
  (a) Proposed experimental configuration for a junction involving two
  cuprate leads A and B, with transition temperatures as indicated by
  the phase diagram of (b).  (c) Diagrammatic representation of the
  incoherent pair tunneling current contribution. Lines ($\pm k,\pm
  k'$) and ( $\pm p, \pm p'$) correspond to single electron
  propagators of the normal pseudogapped (PG) B lead and anomalous
  Gor'kov propagators of the superconducting A lead, respectively.
  The dots ($\bullet$) represent tunneling matrix elements $V_{\bf
    pk}$, and the box stands for the particle-particle t-matrix of B.
  }
\label{fig:1}
\end{figure}

The basic experimental configuration is illustrated in
Fig.~\ref{fig:1}~(a).  An I-V measurement is made on a tunnel junction
formed from an optimally doped cuprate superconductor A and a
non-optimally doped material B, in a temperature range $T_c^A > T >
T_c^B$. The c-axis is perpendicular to the A and B layers which are
separated by an insulating layer. Such a structure could be obtained
by varying the doping concentration of a crystal during a
layer-by-layer deposition \cite{eckstein}.  If the B lead is
underdoped, as indicated in Fig.~\ref{fig:1}~(b), there will be a
substantial temperature region above $T_c^B$ in which B will have a
pseudogap, while if the normal lead is overdoped (denoted by B' in
Fig.~\ref{fig:1}~(b)), this pseudogap region will be significantly
narrower. Now, for $T_c^A > T > T_c^B$ we can use the superconducting
pair field of the optimally doped superconductor to directly probe the
strength of the pairing fluctuations in B, by measuring the incoherent
pair tunneling contribution $I_p(V)$ to the total tunneling current
$I(V)$ \cite{djs,goldman,theory}. If pseudogap behavior is associated
with strong precursor superconducting pairing, the contribution from
the incoherent pair tunneling $I_p(V)$ should extend over a much wider
temperature range than for the overdoped B' lead, even if $T_c^B$ for
the two are equal.  Furthermore, if indeed the pseudogap region is
characterized by precursor pairing, the voltage structure of $I_p(V)$
provides a measure of the frequency dependence of the imaginary part
of the particle-particle t-matrix in the pseudogap regime.

The incoherent pair tunneling contribution to the total tunnel current
$I(V)$ is shown diagrammatically in Fig.~\ref{fig:1}~(c)
\cite{theory}. The quasiparticle states in A are labeled with $p$ and
$p'$ and those in B with $k$ and $k'$: The $p$ lines correspond to the
Gor'kov functions $F_p(i\omega_n)$ of the A lead, while the $k$ lines
represent the single particle Green's functions of the B lead. The
dots represent the one-electron tunneling matrix element $V_{p,k}$ and
$t$ is the particle-particle t-matrix for material B.
It should be stressed that, due to the very short c-axis coherence
length in the cuprates, the pair tunneling takes place between the two
cuprate layers on either side of the insulating barrier.
In order to analyze the experimental requirements, we consider a
circular Fermi surface and assume that the pairing instability
occurs in the d-wave channel with a t-matrix given by

\begin{equation}
t_{\bf k,k',q} (i \omega_m) =  t_{\bf q} (i\omega_m) 
\cos (2\varphi_{\bf k}) \cos (2 \varphi_{\bf k'}).
\end{equation}
Here $i\omega_m = 2m\pi T$ is the bosonic Matsubara frequency (unless
noted otherwise, we take $\hbar = 1, k_B = 1$), and $\varphi_k =
\arctan (k_y/k_x)$.  In the absence of an external magnetic field, the
incoherent pair tunneling contribution $I_p(V)$ is given by $4e$ times
the imaginary part of the diagram shown in Fig.~\ref{fig:1}~(c)
\cite{djs,theory}
\begin{equation}
I_p(V) = 4e C^2 S a^2\, {\rm Im}  \,
t_{\bf q = 0}(i\omega_m \rightarrow 2eV + i\delta),
\label{eq:ip}
\end{equation}
where $S$ is the junction area, $a$ is the lattice spacing, and the
coefficient $C$ -- which determines the magnitude of the pair current -- is
given by the following expression

\begin{eqnarray}
 C &=& \frac{n_i}{N^2}\,T\!\sum_{n,\bf p,k} F_{\bf p} (i\omega_n) G_{\bf k}
  (i\omega_n) G_{\bf -k} (-i\omega_n)   \nonumber\\
 &&\times\left\langle|V_{\bf k,p}|^2\right\rangle_{\text{imp}} \cos (2
  \varphi_{\bf k})\;.
\label{eq:c}
\end{eqnarray}
Here we presume that the mechanism for electron transfer from A to B
derives from impurity assisted hopping in the insulating layer
separating A and B. We define $n_i$ to be the number of impurity
scattering sites per unit area of the insulating layer, $N$ is the
number of sites of a layer, and $\langle |V_{\bf pk} |^2
\rangle_{\text{imp}} $ is the impurity averaged single-electron
transfer. The momenta ${\bf p}$ and ${\bf k}$ are two-dimensional
vectors. In Eq.~(\ref{eq:c}) we have neglected the weak voltage
dependence of $C$ which is justified in the regimes we will be
studying where the t-matrix dominates the voltage dependence of the
pair current.

To estimate the size of the pair current we have used the
Bardeen-Cooper-Schrieffer (BCS) form for the Gor'kov function to
describe the superconducting A lead

\begin{equation}
F_{\bf p} (i\omega_n) = \frac{\Delta_A\cos(2\varphi_{\bf p})}{\omega_n^2 + 
\epsilon^2_{\bf p} + \Delta_A^2 \cos^2(2\varphi_{\bf p})}\;.
\label{eq:f}
\end{equation}
Here $\Delta_A$ is the maximum of the associated d-wave
superconducting gap.  The detailed nature of the Green's functions in
B are of course very important in determining the particle-particle
t-matrix. However, the coefficient $C$ is obtained by summing over
both the momentum and frequency variables of the propagators. Thus $C$
is only marginally affected by the precise form of the single particle
propagators. Whether one replaces the product of the B Green's
functions by their non-interacting form

\begin{equation}
  \label{eq:5}
  G_{\bf k}(i\omega_n) G_{\bf -k}(-i\omega_n) = \frac{1}{\omega_n^2 +
  \epsilon^2_{\bf k}}\;, 
\end{equation}
%
or whether one uses an extreme limit \cite{PG_theories} of
pseudogap theories

\begin{equation}
  \label{eq:6}
  G_{\bf k}(i\omega_n) G_{\bf -k}(-i\omega_n) = \frac{1}{\omega_n^2 +
  \epsilon^2_{\bf k}+\Delta_B^2\cos^2\left(2\varphi_{\bf k}\right)}\;, 
\end{equation}
changes the estimate of $C$ only by factors of order unity
\cite{normanpheno}. In these expressions $\epsilon_{\bf p}$ and
$\epsilon_{\bf k}$ are the single particle energies in A and B,
respectively.

We will assume that the insulating layer gives rise to a diffuse
\cite{hirsh,tunn} electron transfer with

\begin{equation}
\left\langle|\ V_{\bf pk}|^2\right\rangle_{\text{imp}}  =  |V_0|^2 + |V_1|^2
\cos(2\varphi_{\bf p}) \cos(2\varphi_{\bf k}) \;.
\label{eq:V}
\end{equation}
More generally one could imagine expanding the impurity averaged
single-electron transfer $\langle |V_{\bf pk} |^2 \rangle_{\text{imp}}
$ in two dimensional crystal harmonics. In Eq.(\ref{eq:V}) we have
kept only the uniform and d-wave pair transfer parts \cite{norman}. It
is the second term in Eq.(\ref{eq:V}) that will enter in our
calculations. The required size of $V_1$ will be discussed below,
together with other junction requirements.
Using Eqs.~(\ref{eq:c})--(\ref{eq:V}) one finds that

\begin{equation}
  \label{eq:7}
  C = \pi^2 n_i N_A(0) N_B(0) |V_1|^2 \Delta_A T\sum_n I_A(\omega_n)
  I_B(\omega_n)\;, 
\end{equation}
where $N_A(0)$ and $N_B(0)$ are the single-particle density of states per
spin, per site for layer A and B, respectively, and

\begin{eqnarray}
  \label{eq:8}
  I_{A,B}(\omega_n) &=& \int\frac{d\varphi_{\bf p}}{2\pi}
  \frac{\cos^2(2\varphi_{\bf p})}{\sqrt{\omega_n^2+\Delta^2_{A,B} \cos^2 2
  \varphi_{\bf p}}} 
  = \frac{2}{\pi\sqrt{\omega^2_n + \Delta^2_{A,B}}} \nonumber \\
  &&\times \left[E(k_n) +
  \left(\frac{\omega_n}{\Delta_{A,B}}\right)^2 (E(k_n) - K(k_n))\right]\;. 
\end{eqnarray}
In the above expression
$k_n^2=\Delta_{A,B}^2/(\omega_n^2+\Delta_{A,B}^2)$, and $K$ and $E$
are the complete elliptic integrals of the first and second kind,
respectively. Carrying out the Matsubara sum in Eq.~(\ref{eq:7}), we
find that to within numerical factors of order unity

\begin{equation}
  \label{eq:9}
  C \simeq \frac{\pi^2}{4} n_i N_A(0) N_B(0) |V_1|^2 \;. 
\end{equation}
Now at low temperatures, where A and B are both superconducting, a similar
calculation shows that the Josephson critical current is given by

\begin{equation}
  \label{eq:10}
  I_c = 2e C' \Delta_B S\;,
\end{equation}
with $\Delta_B$ the low temperature maximum gap in B and the
coefficient $C'$ is closely related to $C$ given by Eq. (\ref{eq:c})
\cite{DS-note}. Using this to normalize the strength of $I_p(V)$ we
have

\begin{equation}
  \label{eq:11}
  \frac{I_p(V)}{I_c} \approx \frac{E_J}{E_c} \text{Im} \bar{t}(2eV)\;.
\end{equation}
Here $E_J= \hbar I_c/2e$ is the zero temperature Josephson coupling energy
between A and B, $E_c=(S/a^2)N_B(0)\Delta_B^2/2$ is the condensation energy
of the B cuprate layer, and
$\text{Im}\bar{t}(\omega)=N_B(0)\text{Im}t_0(\omega)$ is a dimensionless
form of the t-matrix for ${\bf q}=0$.

It can be seen from Eq.~(\ref{eq:11}) that the important quantity
measured in a pair tunneling experiments is ${\rm Im }{\bar{t}}(\omega)$.
The form of this function varies depending on the particular scenario
adopted for describing the pseudogap.
%
%
For a wide class of theories, ${\rm Im} t_{\bf q} (\omega)$ can be
expressed in terms of the pair susceptibility $\chi_{\bf q}(\omega) $
and the pairing coupling constant $g$

\begin{equation}
  \label{eq:Imt}
  \text{Im}t_{\bf q}(\omega) =
  \frac{-g^2\text{Im}\chi_{\bf q}(\omega)}{[1 +
  g\text{Re}\chi_{\bf q}(\omega)]^2 + 
  [g\text{Im}\chi_{\bf q}(\omega)]^2}  \;.
\end{equation}
A useful form of Eq.~(\ref{eq:Imt}) for experimental comparison is
discussed in Refs.~\cite{Janko,Maly}, although other alternatives may
eventually be proposed using different precursor 
scenarios \cite{2e,2e_a}.
The approach of Refs.~\cite{Janko,Maly} provides a concrete
diagrammatic prescription for computing $\chi$.
%
%
For ${\bf q}=0$ and sufficiently low frequencies,
$1+g\text{Re}\chi_0(\omega)\approx (\alpha/\gamma)(\omega-\omega_o)$ and
$g\text{Im}\chi_0(\omega) \approx \omega/\gamma$, where (for $ T $ close to
$T_c^B$), $\omega_o=(\gamma/\alpha)(T/T_c^B-1)$. The values of $\alpha$ and
$\gamma$ depend on $g$.  Under these conditions, Eq.~(\ref{eq:Imt}) yields

\begin{equation}
\label{eq:Imt1}
{\rm Im} \bar t(\omega ) \simeq \frac{\gamma\omega }{
  \alpha^2(\omega-\omega_o)^2 + \omega^2}\;.
\end{equation}
In the weak-coupling limit, where the dimensionless parameter $\alpha\simeq
0$, Eq.~(\ref{eq:Imt1}) yields the well known result \cite{djs,goldman}

\begin{equation}
  \label{eq:Imt2}
  \text{Im}\bar{t}(\omega) = \frac{\omega/\gamma}{(T/T^B_c-1)^2
  +(\omega/\gamma)^2}\;.
\end{equation}
In this regime the pairing fluctuations are associated with critical
behavior and are essentially diffusive in nature. This case was addressed in
earlier incoherent pair tunneling experiments \cite{goldman} on conventional
superconductors.
By contrast, in the intermediate coupling regime, which corresponds to
$\alpha \simeq 1$, the value of $\omega_o$ is strongly reduced,
resulting in a pronounced \textit{resonance} in
$\text{Im}\bar{t}(\omega)$ at this frequency. Thus, the pair
fluctuations acquire a propagating nature\cite{Janko,Maly}.

\begin{figure}
\centerline{\epsfxsize=2.2in\epsffile{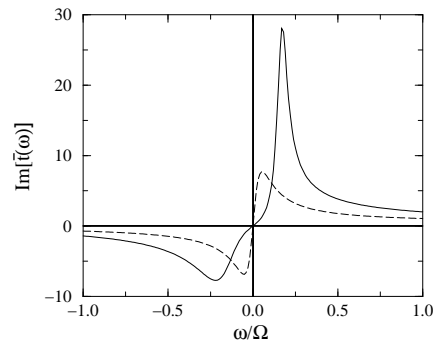}}
\caption{Predicted voltage dependence of the pair tunneling current; 
  following \protect\cite{Maly} the solid
  and dashed curves correspond to under (B) - and overdoped (B') leads
  at $ T/T_c = 1.1$; the slightly doping dependent $\Omega$ is of order
  of several hundred $\text{meV}$.  Dashed curve is similar to that of
  conventional fluctuation picture \protect\cite{djs}.  The asymmetry
  of the solid curve is an important signature which should be noted.}
\label{fig:t}
\end{figure}

These two theoretical limits are illustrated in Fig.~\ref{fig:t} which
presents the self-consistently calculated t-matrix \cite{Maly} in weak
and intermediate coupling, corresponding to $ B'$ and $ B$
respectively.
Here $T/T_c^B = 1.1$. Notice the asymmetry ${\rm Im} \bar t (\omega )
\neq {\rm Im} \bar t (-\omega)$ in the second case which provides a
strong signature for pair resonance effects.  This asymmetry is, in
turn, related to an asymmetric density of states \cite{Janko}, which
may be associated with that observed in STS experiments \cite{renner}.
This figure also reflects the predicted voltage dependence of the pair
tunneling current. Within the superconducting pairing fluctuation
scenarios of the type discussed in Refs.  \cite{Janko,Maly} a
prominent peak in ${\rm Im} \bar t(\omega) $ is expected to persist in
underdoped cuprates to temperatures of order $T \sim T^*$,
considerably higher than $T_c^B$. Alternative scenarios\cite{2e,2e_a}
can be used, presumably, to provide analogous signatures, within their
respective theoretical framework.  The importance of the incoherent
pair tunneling experiment lies in its ability to detect such features
and therefore to confirm or falsify different classes of pseudogap
scenarios.

Let us now estimate the size of the pair current $I_p$ given by
Eq.(\ref{eq:11}).  The condensation energy density can be inferred
from heat capacity measurements \cite{loram}: For an underdoped
$YBa_2Cu_3O_{6+x}$ of $T_c \sim 60 \text{K}$ we obtain $\epsilon_{\rm
  cond} \sim 2 \times 10^4\,\text{J/m}^3$.  Using typical values for
junction surface area $S \sim 10^{-8}\,\text{m}^2$ \cite{Miyakawa} and
taking the layer thickness of order $\ell_c \sim 10\,\text{\AA}$, we
find $E_{\rm c} = \epsilon_{\rm cond} S\ell_c \sim 2 \times
10^{-13}\,\text{J} \approx {10}^6\,\text{eV}$.  Given the
relatively large condensation energy, the strength of the Josephson
coupling becomes crucial for the effect to be observable.
For a typical critical current of order $I_c \sim 10\,\text{mA}$, the
corresponding Josephson coupling energy is $E_{\rm J} = \hbar I_c/2e
\sim 3 \times{10}^{-18}\,\text{J} \approx 20\,\text{eV}$.
Consequently, $I_p \sim I_c(E_J/E_c) \sim 0.2 \,\mu\text{A}$, which is
of the same order as the pair currents detected in conventional
superconductors\cite{goldman}.

Thus it is important to fabricate junctions with c-axis Josephson
current density in the range of $10^2 {\rm A/cm}^2$.  Critical current
densities sustained by recently fabricated trilayer junctions
\cite{eckstein} are in this range.  In order to detect this small
incoherent pair tunneling contribution to the total current I(V), one
must be able to separate it from the larger quasiparticle current
\cite{goldman2}.  Fortunately this can be done by turning on a
magnetic field $\textbf{H}$ in the plane of the junction, as shown in
Fig.~\ref{fig:1}~(a). When $\textbf{H}$ is such as to put several flux
units in the junction, the incoherent pair tunneling will be
suppressed. If the thickness of the A layer is larger than the
in-plane penetration depth, $\lambda^A_{ab}(T)$, and the B layer is
thin compared to this, then one flux unit will be present when
$\text{H} L \lambda_{ab}(T) \approx hc/(2e)$. For $L = 100\,\mu\text{m}$ and
$\lambda^A_{ab} \sim 0.2\,\mu\text{m}$ and $d \sim 500 \text{\AA}$, one has
$\text{H} \sim 0.1$ Gauss. In order to observe the zero field pair
current $I_p(V)$, it will be necessary to magnetically shield the
junction.
Then by subtracting the I--V data in the presence of an external
magnetic field $\text{H}\approx 1\,\text{Gauss}$ from the I--V data in
the absence of $\text{H}$ one can obtain the incoherent pair tunneling
contribution $I_p(V)$.
Further complications might be caused by thermal voltage noise
\cite{goldman,kadin} in the junction circuit due to the relatively
elevated temperatures at which these measurements need to be carried
out. One possibility would be to use the single-layer Bi2201 compound
for both leads: this material has a phase diagram similar to that in
Fig.  \ref{fig:1}~(b), but with a relatively low optimal $T_c$.

In conclusion, we have argued that the measurement of the pair
tunneling current between an optimally doped and an underdoped cuprate
can be used to probe the pairing fluctuations in the pseudogap state.
This experiment has, in principle, the potential to reveal whether the
pseudogap state is in fact due to pairing fluctuations. Indeed, strong
pairing correlations in the pseudogap state will be manifest in a
large pair current, as compared to the pair current of a junction
where an overdoped lead of the same $T_c$ is used.  No such strong
doping dependence of the pair current is expected within pseudogap
scenarios that do not invoke the onset of strong pairing correlations
below $T^*$. To illustrate this experiment, we have chosen the
particular case of a c-axis junction geometry and identified the
region of the phase diagram where the experiment should be performed.
We have also established the range of basic junction parameters
suitable for observing the pair current, and suggested an experimental
procedure for separating the small pair contribution from the large
quasiparticle current.

We would like to thank I.~Bozovic, Q.~Chen, A.M.~Goldman, D. van
Harlingen and J.F. Zasadzinski for useful discussions, and to the
Aspen Center for Physicsfor their hospitality during the HighTc
Summer Workshop 1998, where part of this research was performed.
This research was supported in part by the National Science Foundation
under awards DMR91-20000 (administrated through the Science and
Technology Center for Superconductivity; B.J., I.K., K.L.),
DMR95-27304 (D.J.S.), and the U.S.  Department of Energy, Basic Energy
Sciences, under Contract No.~W-31-109-ENG-38 (M.N.).

\begin{references}
\bibitem{PG_theories}
The Green's functions in theories such as those of Refs.~\cite{Janko,Maly} are
  more complicated than those in Eq.~(\ref{eq:6}), and
  contain additional parameters besides the pseudogap $\Delta_B$. However,
  these complications have little influence on the actual size of $C$.
\bibitem{DS-note}
In Eq.~(\ref{eq:10}), $C'$ is the low temperature limit of Eq.~(\ref{eq:7})
  which again gives a value for $C$ which is to within factors of order
  $\Delta_A/(\Delta_A+\Delta_B)$ the same as Eq.~(\ref{eq:9}).
\end{references}
\end{document}